\documentclass[11pt,a4paper]{article}
\pdfoutput=1

\usepackage{jcappub}
\usepackage{psfrag,color,epsfig}
\usepackage{graphicx,graphics}	
\usepackage{amsmath}	
\usepackage{amssymb}	
\usepackage{multicol}	
\usepackage{bm}		
\usepackage{pdflscape}	
\usepackage{multirow}
\usepackage{tabularx}  
\usepackage{physics}
\usepackage{xspace}

\usepackage[T1]{fontenc}
\usepackage{ae,aecompl}
\usepackage{newtxmath}
\usepackage{ulem}
\def\HI{{H}{\scriptsize I}~}
\def\kk{{\vec{k}}}

\def\x{\vec{x}}
\def\mpc{{\rm Mpc}}
\def\impc{{\rm Mpc}^{-1}}
\def\Nt{N_{\rm tri}}

\title{A fast estimator for quantifying the shape dependence of the 3D bispectrum}
\author[a,1]{Abinash Kumar Shaw,\note{Corresponding author.}}
\author[a]{Somnath Bharadwaj,}
\author[b]{Debanjan Sarkar,}
\author[a]{Arindam Mazumdar,}
\author[a]{Sukhdeep Singh}
\author[c,d]{and Suman Majumdar}

\affiliation[a]{Department of Physics \& Centre for Theoretical Studies, Indian 
Institute of Technology Kharagpur,\\Kharagpur 721302, India}
\affiliation[b]{Department of Physics, Ben-Gurion University of the Negev, \\ Be'er Sheva 84105, Israel}
\affiliation[c]{Department of Astronomy, Astrophysics and Space Engineering, Indian Institute of Technology Indore,\\ Simrol, Indore 453552, India}
\affiliation[d]{Department of Physics, Blackett Laboratory, Imperial College, \\ London SW7 2AZ, U. K.}

\emailAdd{abinashkumarshaw@gmail.com}
\emailAdd{somnath@phy.iitkgp.ac.in}

\date{\today}

\abstract{
The dependence of the bispectrum on the size and shape of the triangle contains a wealth of cosmological information. Here we consider a triangle parameterization which allows us to separate the size and shape dependence. We have implemented an FFT based fast estimator for the three dimensional (3D) bin averaged bispectrum, and we demonstrate that it allows us to study the variation of the bispectrum across triangles of all possible shapes (and also sizes). The computational requirement is shown to scale as $\sim N_{\rm g}^3 \, \log{N_{\rm g}^3}$ where $N_{\rm g}$ is the number of grid points along each side of the  volume. We have validated the estimator using a non-Gaussian  field  for which the bispectrum can be analytically calculated. The estimated bispectrum values are found to be in good agreement ($< 10 \%$ deviation) with the analytical predictions across much of the triangle-shape parameter space. We also introduce linear redshift space distortion, a situation where also the bispectrum can be analytically calculated. Here the estimated bispectrum is found to be in close agreement with the analytical prediction for the monopole of the redshift space bispectrum.
}

\arxivnumber{2107.14564}

\begin{document}
\maketitle
\flushbottom


\section{Introduction}
\label{sec:intro}

The simplest inflationary models predict the primordial large-scale fluctuations to be a Gaussian random field \cite{Baumann_2009} for which the power spectrum is adequate to characterize the statistical properties fully. However, the matter density fluctuations are predicted to become non-Gaussian as they evolve \cite{Fry_1984} due to the non-linear growth and the non-linear biasing. Further, another class of inflationary models predicts the primordial fluctuations to be non-Gaussian \cite{Bartolo_2004}. Whatever be the case, it is necessary to consider higher-order statistics for a complete  description of the large-scale structures in the Universe. The three-point correlation function (3PCF) or its Fourier conjugate, the bispectrum, is the lowest order statistics which is sensitive to non-Gaussianity. Measurements of the bispectrum from observations of the Cosmic Microwave Background (CMB) (e.g. \cite{Sefusatti_2009,Fergusson_2012,Oppizzi_2018, Planck_NG_2018,Shiraishi_2019}) and the galaxy surveys (e.g. \cite{Feldman_2001, Scoccimarro_2004, Liguori_2010, Scoccimarro_2015, Ballardini_2019, Pearson_2018}) have been used to place stringent constraints on the primordial non-Gaussianity. Second-order perturbation theory ($2$PT) predicts \cite{Matarrese_1997} that the bispectrum measurements in a quasi-linear regime can be used to determine the bias parameters, and following this the galaxy bias parameters have been quantified from different galaxy surveys (e.g. \cite{Feldman_2001, Scoccimarro_2001, Verde_2002, Nishimichi_2007, Sefusatti_2007, Gil-Marin_2016}). Further, the measurements of bispectrum enable us to lift the degeneracy between $\Omega_m$ (appearing in $f(\Omega_m)$) and $b_1$, something that is not possible by considering only the power spectrum \cite{Scoccimarro_1999}.

Redshift space distortion is important in the context of the bispectrum, and this has been extensively studied in the literature (e.g. \cite{Hivon_1995, Verde_1998, Scoccimarro_1999, Hashimoto_2017, Nan_2018, Desjacques_2018}). Recently \cite{Yankelevich_2018} and \cite{Gualdi_2020} have presented predictions for cosmological parameter estimation considering the redshift space bispectrum and power spectrum. \cite{Clarkson_2019} and \cite{de_Weerd_2020} have recently shown that relativistic effects will introduce a dipole anisotropy in the redshift space bispectrum on very large length-scales.

Ongoing and upcoming future galaxy surveys like DESI\footnote{\href{https://www.desi.lbl.gov/}{https://www.desi.lbl.gov/}} \cite{Levi_2013}, LSST\footnote{\href{https://www.lsst.org/}{https://www.lsst.org/}} \cite{Ivezic_2019}, EUCLID\footnote{\href{https://www.euclid-ec.org/}{https://www.euclid-ec.org/}} \cite{Laureijs_2011} are aimed to cover large volumes of the order of several ${\rm Gpc}^3$ in the sky. It is essential to develop fast and accurate bispectrum estimators for the analysis of these large galaxy surveys. \cite{Slepian_2017} and \cite{Slepian_2018} present a fast technique to quantify the redshift space 3PCF function by expanding it in terms of products of two spherical harmonics. \cite{Sugiyama_2018} have proposed a tri-polar spherical harmonic decomposition to quantify the anisotropy of the redshift space bispectrum. They have also demonstrated this by applying it to the Baryon Oscillation Spectroscopic Survey (BOSS) Data Release 12. Apart from this, the upcoming SKA\footnote{\href{https://www.skatelescope.org/}{https://www.skatelescope.org/}} (Low \& Mid, \cite{Huynh_2013}) will map large volumes of the universe at a high level of sensitivity using the redshifted 21-cm radiation. Measurements of the bispectrum of the 21-cm radiation from the cosmic dawn \cite{Watkinson_2018, Kamran_2021}, the epoch of reionization (e.g. \cite{Bharadwaj_2005, Shimabukuro_2017, Majumdar_2018, Majumdar_2020, Watkinson_2021, Hutter_2019, Trott_2019, Saxena_2020}) and the post-reionization era (e.g. \cite{Ali_2006, Sarkar_2019, Durrer_2020, Cunnington_2021}) are expected to yield a wealth of information regarding the evolutionary history of the universe.

In a recent work \cite{Bharadwaj_2020} (and also \cite{Mazumdar_2020}) have quantified the effect of redshift space distortion on the bispectrum. In general, the bispectrum depends on the shape and size of the triangle formed by three Fourier modes. The above-mentioned work presents a very convenient method to parameterize the shape and size of a triangle using the length of the largest side $k_1$ for the size and two dimensionless parameters $\mu,t$ for the shape.  Each set of  $(k_1,\mu,t)$ represent a different triangle, and the allowed range of the parameter values uniquely covers all possible shapes and sizes with no repetition. In the present paper, we have implemented a fast estimator for evaluating the binned bispectrum across the parameter space $(k_1,\mu,t)$. The estimator is validated using a non-Gaussian density field for which the expected bispectrum is known. Nearly most of the earlier works which have implemented and validated bispectrum estimators (e.g. \cite{Sefusatti_2016, Schmittfull_2013, Watkinson_2017}) have restricted the analysis to triangles with a very limited range of shapes such as isosceles and equilateral triangles. In addition to these, few works (e.g. \cite{Gil-Marin_2016, Byun_2017, Byun_2021}) have considered different triangle configurations (isosceles, equilateral and scalene) in their implementations. However the index representation of triangle configurations are not so intuitive for interpreting any measured bispectrum. In contrast, the analysis presented here spans the entire $(\mu,t)$ space covering triangles of all possible unique shapes. The parameterization used here is more convenient and intuitive for interpreting the bispectrum. We further provide a visual representation of the shape dependence of the bispectrum covering triangles of all possible shapes. It is important to note that the shape dependence contains significant cosmological information which is important for interpreting the bispectrum \cite{Matarrese_1997}.   

The fast bispectrum estimator implemented here is based on the Fourier transform (FT) bispectrum estimation technique whose mathematical framework has been presented in the theses of \cite{Sefusatti_thesis} and \cite{Jeong_thesis}. FT based estimators have  been used in several earlier works (e.g. \cite{Scoccimarro_2015, Gil-Marin_2016, Pearson_2018}) to estimate the  bispectrum for various galaxy  surveys, and also to constrain primordial non-Gaussianity from the CMB bispectrum \cite{Smith_2013}. Recently, \cite{Watkinson_2017, Watkinson_2018, Watkinson_2021} have applied an FT based estimator to estimate the 21-cm bispectrum  from the epoch of reionization simulations. As mentioned earlier, the parameterization and binning adopted in the present paper is different from earlier works. This parameterization helps to elucidate the shape dependence and visually represent this. The binning adopted here naturally spans the space of triangles of all possible shapes. In this paper, we briefly present the mathematical framework for the estimator and validate the estimator using simulations in real space. Further, we have also validated our estimator in the presence of redshift space distortion which introduces an anisotropy along the line-of-sight (LoS) direction. A brief outline of the paper follows.

We present the mathematical formulation of the FFT based Bispectrum Estimator (FBE) and the methodology in Section \ref{sec:meth}. Here we also demonstrate the improvement of the FBE over the Direct Bispectrum Estimator (DBE) in terms of computation time. Next, we present the validation of our estimator in Section \ref{sec:valid} using simulated non-Gaussian field. Finally, we summarize and discuss our findings in Section \ref{sec:dis}. 

\section{Methodology}
\label{sec:meth}

\subsection{Binned Bispectrum Estimator}
\label{subsec:estim}
\begin{figure}
    \centering
    \resizebox{.45\textwidth}{!}{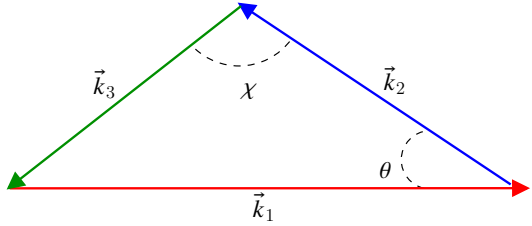}
    \caption{Shows a closed triangle $\kk_1+\kk_2+\kk_3=0$ where $\theta$ is the inner angle between $\kk_1$ and $\kk_2$, and $\chi$ is the inner angle between $\kk_2$ and $\kk_3$. We use $\mu=\cos\theta$ to parameterize the shape of triangles.}
    \label{fig:tri}
\end{figure}

The bispectrum $B(k_1,k_2,k_3)$ of any random field $\delta(\x)$ within a finite region of volume $V$ is defined through 
\begin{equation}
\delta_{\rm K}(\kk_1+\kk_2+\kk_3)~  B(k_1,k_2,k_3)=V^{-1} \langle \Delta(\kk_1) \Delta(\kk_2) \Delta(\kk_3) \rangle,
\label{eq:bs}
\end{equation}
where $\Delta(\kk)$ is the Fourier transform of $\delta(\x)$ and $\langle \cdots \rangle$ denotes the average with respect to an ensemble of independent realizations of the random field $\delta(\x)$. Here we assume that $\delta(\x)$ is statistically homogeneous and isotropic, a valid assumption in the absence of any LoS anisotropy in the signal e.g. redshift space distortions. The Kronecker delta $\delta_{\rm K}(\kk_1+\kk_2+\kk_3)$ in the above equation indicates that for a statistically homogeneous random field the bispectrum is only defined when the three wave vectors $\kk_1$, $\kk_2$ and $\kk_3$ form a closed triangle as shown in Fig. \ref{fig:tri}.

In the absence of redshift space distortion or any such LoS anisotropy, the bispectrum depends only on the triangle formed by the three vectors $(\kk_1,\kk_2,\kk_3)$, and is independent of the orientation of the triangle. A triangle can be completely specified using the lengths of the three sides namely $(k_1, k_2, k_3)$ where $k_a=|\kk_a|$, and we can parameterize the bispectrum using $B(k_1, k_2, k_3)$. The value of the bispectrum depends on both the shape and the size of the triangle $(k_1, k_2, k_3)$, however, this particular parameterization does not allow us to separate these two. In this paper, we adopt a parameterization \cite{Bharadwaj_2020} which allows us to separately quantify the shape and the size dependence of the bispectrum.

\begin{figure}
    \centering
    \includegraphics[scale=0.6]{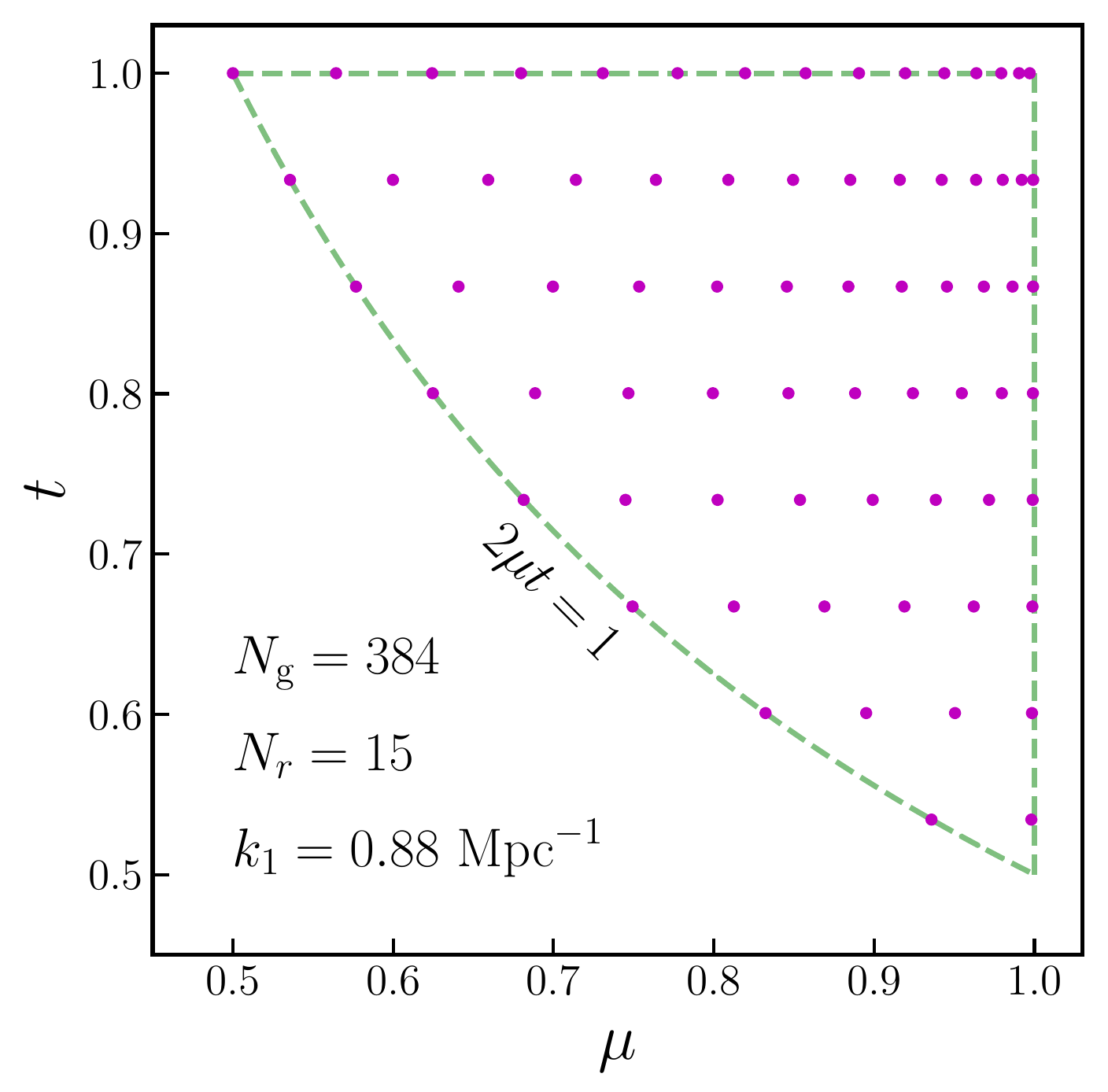}
    \caption{Shows the allowed regime of $(\mu,t)$ space  which uniquely covers triangle of all possible shapes. The boundaries $t=1$ and $2\mu t=1$ correspond to L and S isosceles triangles respectively, whereas the boundary $\mu=1$ corresponds to linear triangles. The magenta points show the sampling of $(\mu,t)$ space which is obtained considering $N_{r}=15$ annular rings of equal width $\delta k=0.0584~\impc$. The value of $k_1$ here is fixed at 
    $k_1=0.88~\impc$ which corresponds to the center of the largest $(15^{\rm th})$ annular ring.}  
    \label{fig:sampling}
\end{figure}

We proceed by labelling the three sides of the triangle such that $k_1\geq k_2 \geq k_3$. Following \cite{Bharadwaj_2020}, we use the length of the largest side $k_1$ to parameterize the size of the triangles. We parameterize the shape of the triangles (Fig. \ref{fig:tri}) using $\mu=\cos{\theta}= -(\kk_1\cdot \kk_2)/(k_1~k_2)$ which is the cosine of the angle between $-\kk_2$ and $\kk_1$, and $t=k_2/k_1$ which is the ratio of the second largest side to the largest side. The values of $\mu$ and $t$ are restricted to the range 
\begin{equation}
0.5 \le t,\, \mu \le 1 \quad {\rm and}\quad 2 \mu t \ge 1 \, ,
\label{eq:cond}
\end{equation}
and these uniquely specify the shapes of all possible triangles. Fig. \ref{fig:sampling} shows the allowed range of the parameters $(\mu,t)$. Here the right boundary $\mu=1$ corresponds to linear triangles where $\kk_1$, $-\kk_2$ and $-\kk_3$ are parallel (Fig. \ref{fig:tri}). The top right corner $(\mu \rightarrow 1, t \rightarrow 1)$ and 
and the bottom right corner $(\mu \rightarrow 1, t \rightarrow 0.5)$ correspond to squeezed $(\kk_1=-\kk_2, \kk_3 \rightarrow 0)$ and stretched $(\kk_2=\kk_3 = -\kk_1/2)$ triangles respectively. The top boundary $t=1$ corresponds to L-isosceles triangles where the two larger sides ($\kk_1$ and $\kk_2$) are of equal length, whereas the bottom boundary $2 \mu t =1$ (dashed line) corresponds to S-isosceles triangles where the two smaller sides ($\kk_2$ and $\kk_3$) are of equal length. The top left corner $(\mu \rightarrow 0.5, t \rightarrow 1)$ corresponds to equilateral triangles. The diagonal line $\mu=t$ corresponds to right-angle triangles ($\chi=90^{\circ}$ in Fig. \ref{fig:tri}) while the upper $(t > \mu)$ and lower $(t < \mu)$ halves correspond to the acute and obtuse triangles respectively. The reader is referred to Fig.~2 of \cite{Bharadwaj_2020} for further details. In the subsequent discussion, we use $(k_1,\mu,t)$ to parameterize all possible triangles, and we denote the bispectrum as $B(k_1,\mu,t)$.

\begin{figure}
    \centering
    \resizebox{.6\textwidth}{!}{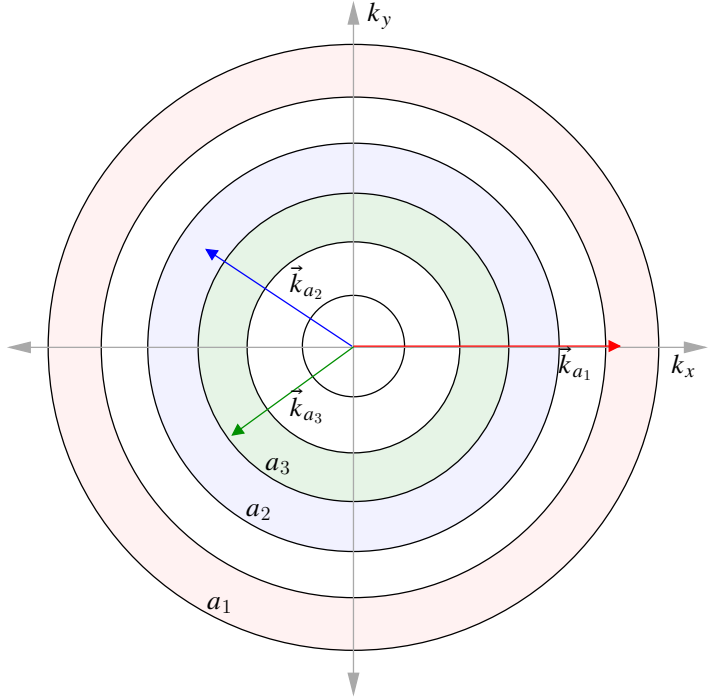}
    \caption{Shows how the two dimensional $\kk$ space is divided into several annular rings. A combination of three rings (labeled $a_1,a_2,a_3$ here) defines a single bin of triangles. A combination of three vectors $(\kk_{a_1},\kk_{a_2},\kk_{a_3})$ which form a closed triangle is illustrated here. The bin contains all such triangles.}
    \label{fig:bins}
\end{figure}

Let us now consider estimating the bispectrum using a cubic volume with $N_{\rm g}$ grid points along each side,  for which the total number of triangles is of the order of $N_{\rm g}^{6}$. Even for a very modest number like $N_{\rm g}\sim 100$ we expect bispectrum estimates for $\Nt \sim 10^{12}$ triangles which is a considerably large number. To reduce the data volume and also the statistical fluctuations, we bin the triangles and typically consider the average bispectrum in $\sim 10$ to a few hundred bins. The binning scheme adopted here is demonstrated in  Fig.~\ref{fig:bins} where we have considered a two dimensional (2D) situation for the sake of simplicity. We have divided the $\kk$ space into $N_r$ annular rings, the  rings  being labelled as $a_1,~a_2,~a_3,...$.  Considering any ring  $a_n$, we use $\kk_{a_n}$ to denote the various $\kk$ modes within the ring. Further $k_n$ and $\delta k_n$ respectively denote the average length of the modes $\kk_{a_n}$ and the radial extent of the ring. 

It is necessary to consider a combination of three annular rings in order to define triangles for estimating the bispectrum, and Fig.~\ref{fig:bins} illustrates one such combination. In keeping with our parameterization, we label the three rings as $a_1,~a_2$ and $a_3$ such that $k_1 \geq k_2 \geq k_3$. We define the binned bispectrum estimator as 
\begin{equation}
\Hat{B}(k_1,k_2,k_3)=\frac{1}{V~\Nt}\sum_{\kk_{a_1}} \sum_{\kk_{a_2}} \sum_{\kk_{a_3}} \Delta(\kk_{a_1}) \Delta(\kk_{a_2}) \Delta(\kk_{a_3}) \,\delta_{\rm K}(\kk_{a_1}+\kk_{a_2}+\kk_{a_3})~,
\label{eq:binbs}
\end{equation}
where the Kronecker delta $\delta_{\rm K}(\kk_{a_1}+\kk_{a_2}+\kk_{a_3})$ enforces that the estimator picks up a contribution only when the three vectors $(\kk_{a_1},\kk_{a_2},\kk_{a_3})$ form a closed triangle. One such triangle is shown in Fig~\ref{fig:bins}. The particular bin shown in this figure contains all possible triangles which have one $\kk$ vector in each of the three respective rings, and $\Nt$ in eq.~(\ref{eq:binbs}) refers to the total number of such triangles in the bin. We expect all the triangles in this bin to have nearly similar size and shape provided the $\delta k$ is sufficiently small for all the three rings. We label each bin using $(k_1,\mu,t)$ where $k_1$ quantifies the average size of the triangles in the bin, and $(\mu,t)$ which are calculated using  
\begin{equation}
\begin{split}
    t &=\frac{k_2}{k_1} \\
    \mu &=\frac{1}{2} \left[\frac{k_1}{k_2}+ \frac{k_2}{k_1} - \frac{k_3}{k_1} \frac{k_3}{k_2}\right]~,
\end{split}
\label{eq:shape}
\end{equation}
quantifies the average shape of the triangles in the bin. Considering the ensemble average of the estimator we can express this as 
\begin{equation}
\langle \hat{B}(k_1,k_2,k_3) \rangle = \frac{1}{\Nt} \sum_{\kk_{a_1}} \sum_{\kk_{a_2}} \sum_{\kk_{a_3}} B(k_1,k_2,k_3) \, \delta_{\rm K}(\kk_{a_1}+\kk_{a_2}+\kk_{a_3})~.
\label{eq:binavgbs}
\end{equation}
We see that the estimator eq.~(\ref{eq:binbs}) measures the binned bispectrum for a bin where the average triangle has size and shape $k_1$ and $(\mu,t)$ respectively. In the subsequent discussion we use the notation $ \langle \hat{B}(k_1,\mu,t) \rangle \equiv \langle \hat{B}(k_1,k_2,k_3) \rangle$ and also do not explicitly show the angular brackets unless necessary to avoid confusion.

In order to illustrate the binning of the triangles, we have considered a 2D box of comoving length $L=215~\mpc$ and $N_{\rm g}=384$ along each side. We consider $N_r=15$ annular rings in the conjugate $\kk$ space. The rings all have  uniform $\delta k=0.0584~\impc$ which corresponds to $2$ grid spacing, and we have $k=0.88~\impc$ for the largest ring. There are ${{}^{17}{\rm C}_3}=680$ possible combinations of three rings, and each such combination corresponds to a distinct bin of triangles for which we may obtain an estimate of the bispectrum. There exists some combinations of rings which do not form any closed triangle. Restricting our attention to the bins with $k_1=0.88~\impc$, we have  ${{}^{16}{\rm C}_2}=120$ possible combinations of annular rings each of which corresponds to a different set of $(\mu,t)$ values. Fig.~\ref{fig:sampling} shows the sampling of $(\mu,t)$ space obtained from this binning procedure for $k_1=0.88~\impc$. We see that the entire $(\mu,t)$ space, which covers triangles of all possible shapes, is quite well covered by this binning procedure. However, we note that the sampling is not uniform and the sampling density increases as we approach $\mu=1$ which corresponds to linear triangles. The number of bins (and sampling) goes down if we consider a smaller $k_1$ value, and for example we have ${{}^{15}{\rm C}_2}=105$ possible bins if we consider $k_1=0.82~\impc$ which corresponds to the second largest annular ring. It is possible to increase the number of bins by increasing $N_r$, however this is limited by the fact that it is not very meaningful to reduce $\delta k$ below a single grid size. Further, the number of triangles in the individual bins goes down if we reduce $\delta k$ and this enhances the cosmic variance in the estimated bispectrum. It is therefore necessary to judiciously choose the value of $\delta k$ so as to optimize two competing factors namely the cosmic variance and the sampling of $(\mu,t)$ space which is crucial to discern the shape dependence of the bispectrum. The annular rings are replaced with spherical shells if we consider a three  dimensional (3D) volume  instead of a 2D area. Retaining the same boundaries as those in 2D, the values of $(k_1,k_2,k_3)$ are slightly different for 3D, however the sampling of $(\mu,t)$ is very similar to that shown in Fig.~\ref{fig:sampling}. 

The choice of the rings (shells) which decides how the triangles are binned is an important factor for our estimator. We have also considered a situation where we have rings of equal logarithmic spacing for which the width $\delta k$ increases with $k$. In this case we find that many of the bins, particularly those which include rings with large $\delta k$, have triangles whose shapes span across a relatively large range of $(\mu,t)$ values i.e. the assumption that all the triangles in any bin have nearly the same shape breaks down. Based on this, we advocate that it is preferred to use rings (shells) of a relatively small, fixed thickness $\delta k$. It may be noted that we still have the freedom of increasing the signal-to-noise ratio (SNR) for the estimated bispectrum by merging the adjacent bins in regions of $(\mu,t)$ space which are more densely sampled than others (e.g. near $\mu=1$ in Fig.~\ref{fig:sampling}). Further, it may also be possible to enhance the SNR by merging the bins which have slightly different values of $k_1$ but the same (or very close) values of $(\mu,t)$. However, for the present analysis we have just used the bins obtained from unique triplets of linearly-spaced rings (shells) and not tried to re-bin the data to increase the SNR.

We now discuss the computation involved in evaluating the bispectrum estimator (eq.~\ref{eq:binbs}). Considering a straightforward and direct implementation, in the first step we  loop through all the  $\kk$ modes to identify the modes in each annular ring (spherical shell in 3D) which involves  $N_{\rm g}^2$ ($N_{\rm g}^3$) steps. Considering a particular combination of three rings (shells)  which respectively  contain $N_{a_1}$, $N_{a_2}$   and $N_{a_3}$ number of $\kk$ modes, we need $N_{a_2} \times N_{a_3}$  steps to loop through all possible combinations of $\kk_{a_2}$ and $\kk_{a_3}$. Only those combinations where $\kk_{a_1}=-(\kk_{a_2}+\kk_{a_3})$ lies within the boundaries of the ring (shell) $a_1$ will contribute to the bispectrum estimator  (eq.~\ref{eq:binbs}). For example,  considering   $k_1=0.88~\impc$ with  $(\mu=0.5,\,t=1)$ which corresponds to equilateral triangles we have $N_{a_2}=N_{a_3}=3280 (776284)$ which implies $10758400 (602616848656)$ steps in the above example if we consider 2D (3D). The number of steps scales as $N_{\rm g}^4$ ($N_{\rm g}^6$) if we increase the number of grid points. This increase is particularly steep in 3D where we have a $64$ fold increase in the number of steps if $N_{\rm g}$ is doubled, and this factor is $729$ if $N_{\rm g}$ is tripled. Further, we need to consider $\sim 100$ such bispectrum estimates in order to quantify the shape dependence for a fixed $k_1$, and we need to repeat this for different $k_1$ for the size dependence. It is quite obvious that the Direct Bispectrum Estimator (DBE) outlined here is computationally extremely demanding, and it is desirable to have a more efficient implementation of the bispectrum estimator (eq.~\ref{eq:binbs}). The Fast Bispectrum Estimator (FBE) described in the next section presents a computationally  efficient implementation of the bispectrum estimator.

\subsection{Fast Bispectrum Estimator}
\label{subsec:fbe}
Considering the bispectrum estimator (eq. \ref{eq:binbs}), we can represent the 3D Kronecker delta as a summation of plane waves over grid points $\x$ in real space as 
\begin{equation}
    \delta_{\rm K}(\kk_{a_1}+\kk_{a_2}+\kk_{a_3}) = \frac{1}{N_{\rm g}^3} \sum_{\x} \exp(-i [\kk_{a_1}+\kk_{a_2}+\kk_{a_3}]\cdot\x)~.
    \label{eq:kdelta}
\end{equation}
It is now possible to use this to express the estimator as 
\begin{equation}
    \hat{B}(k_1,k_2,k_3) =\frac{1}{V~\Nt}\frac{1}{N_{\rm g}^3} \sum_{\x} D(k_1,\x) D(k_2,\x) D(k_3,\x) ~,
\label{eq:bsfin}
\end{equation}
where 
\begin{equation}
    D(k_n, \x)=\sum_{\kk_{a_n}} \Delta(\kk_{a_n})  \exp(-i\kk_{a_n} \cdot\x)~. 
    \label{eq:iFT}
\end{equation} 
We now consider $\Nt$ which appears in eq.~(\ref{eq:bsfin}). The total number of triangles in the particular bin 
\begin{equation}
\Nt = \sum_{\kk_{a_1}} \sum_{\kk_{a_2}} \sum_{\kk_{a_3}} \delta_{\rm K}(\kk_{a_1}+\kk_{a_2}+\kk_{a_3})~.
\label{eq:Ntri1}
\end{equation}
can be calculated using 
\begin{equation}
    \Nt =\frac{1}{N_{\rm g}^3}  \sum_{\x} I(k_1,\x) I(k_2,\x) I(k_3,\x)~, 
\label{eq:Ntri2}
\end{equation}
where 
\begin{equation}
    I(k_n, \x)=\sum_{\kk_{a_n}}  \exp(-i\kk_{a_n} \cdot\x)~. 
    \label{eq:2FT}
\end{equation}

We see that it is now possible to independently carry out the three $\kk$ sums in eqs.~(\ref{eq:binbs}) and (\ref{eq:Ntri1}) without having to explicitly check whether the three modes form a closed triangle or not.  We have evaluated eqs.~(\ref{eq:binbs}) and (\ref{eq:Ntri1}) using the Fast Fourier Transform  (FFT). This, as we discuss later, greatly reduces the computation. The various $D(k_n,\x)$ and $I(k_n,\x)$, once calculated for all the shells, can be saved and utilized to estimate the bispectrum for different bins.

\begin{figure}
    \centering
    \includegraphics[scale=0.45]{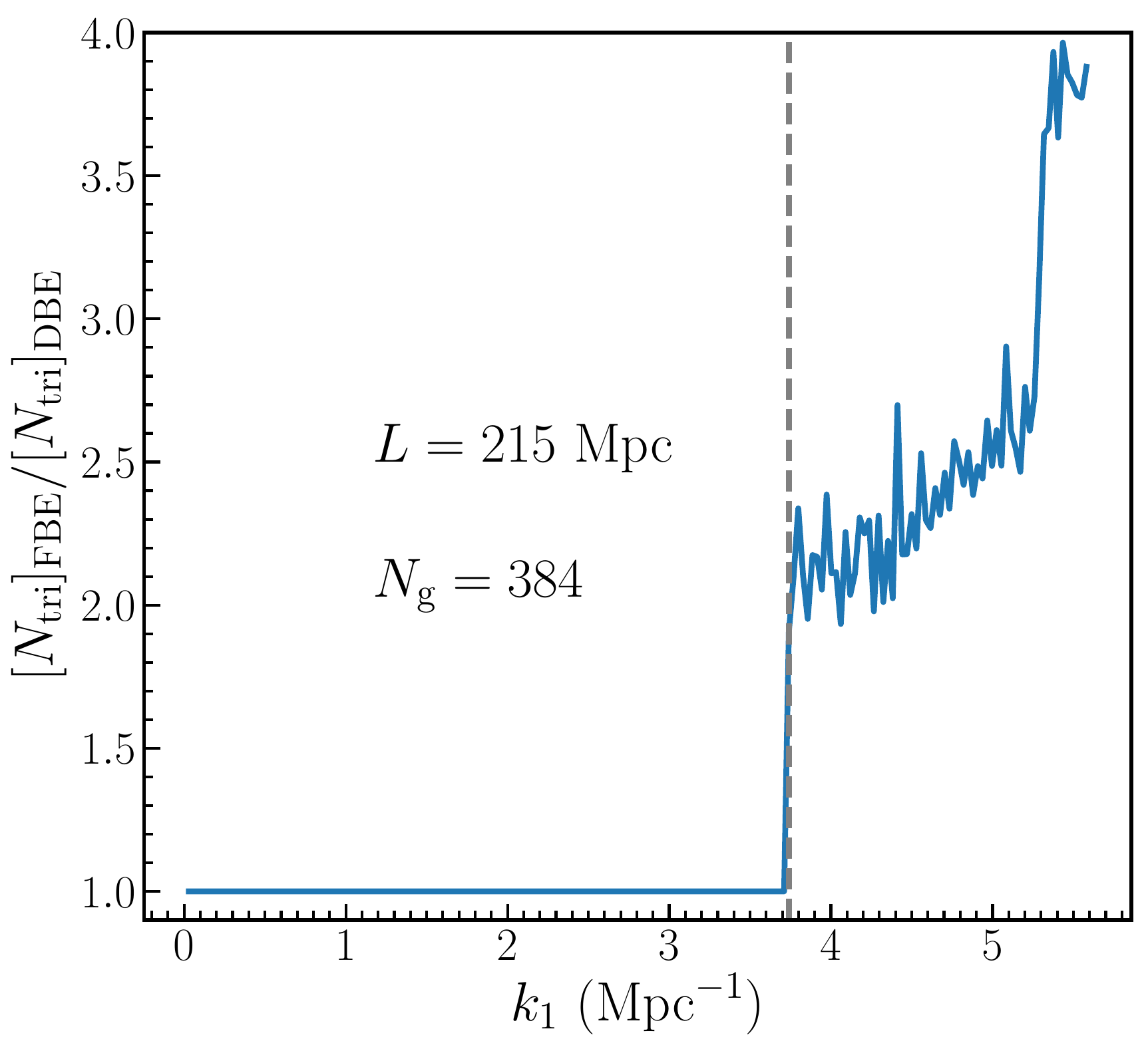}
    \caption{Shows the ratio of the number of triangles obtained using the FBE (eq.~\ref{eq:Ntri2}) 
    to that obtained from the DBE (eq.~\ref{eq:Ntri1}) as a function of $k_1(=k_2=k_3)$ for the $(\mu,t)$ bin corresponding to equilateral triangles. The dashed vertical line demarcates the value  $k_1=(2\pi N_{\rm g})/(3 L)=3.74~\impc$ beyond which the ratio deviates from unity.}
    \label{fig:Nrat}
\end{figure}

The fact that we have a finite volume with periodic boundary conditions imposes a restriction on the FBE. To appreciate this we compare $\Nt$ calculated using eq.~(\ref{eq:Ntri2}) in the FBE with that obtained using eq.~(\ref{eq:Ntri1}) (DBE) which explicitly loops through all possible combinations of $\kk$ values and counts the number of triangles satisfying $\kk_{a_1}+\kk_{a_2}+\kk_{a_3}=0$. Fig. \ref{fig:Nrat} shows $[\Nt]_{\rm FBE}/[\Nt]_{\rm DBE}$ which is the ratio of $\Nt$ calculated using these two methods as a function of $k_1 \, (=k_2=k_3)$ considering rings with $\delta k=0.0292~\impc$ (single grid spacing) for the 2D box described earlier. We see that for $k_1 < 3.74~\impc$ the ratio is $1$ as expected. However, the ratio suddenly shoots up for $k_1 \geq 3.74~\impc$ where we find that FBE predicts a larger number of triangles. To understand this, note that the r.h.s. of eq.~(\ref{eq:Ntri2}) is non-zero for any combination of $(\kk_{a_1},\kk_{a_2},\kk_{a_3})$ which satisfies 
\begin{equation}
  \kk_{a_1}+\kk_{a_2}+\kk_{a_3}= \vec{q} \left( \frac{2 \pi N_{\rm g}}{L} \right) 
\label{eq:3FT}
\end{equation}
where $\vec{q}$ is a vector whose components $(q_x,q_y)$ (and $q_z$ in 3D) are integers. Here the non-zero values of $\vec{q}$ correspond to a situation where on starting from the origin we again return to the origin  upon successively traversing the displacements $\kk_{a_1},\kk_{a_2}$ and $\kk_{a_3}$, however the path winds around the box at least once if not more. While these vectors $(\kk_{a_1},\kk_{a_2},\kk_{a_3})$ also form closed triangles in a periodic box, we cannot interpret their shape and size on the same footing as the ones which satisfy $\kk_{a_1}+\kk_{a_2}+\kk_{a_3}=0$. The discrepancy seen in Fig. \ref{fig:Nrat} arises due to the contribution from the non-zero $\vec{q}$ values. This is borne out by the fact that the discrepancy is only seen when $k_1$ exceeds $ (2 \pi N_{\rm g})/(3 L)= 3.74~\impc$. In the subsequent application of the FBE we avoid the contribution from non-zero $\vec{q}$ values be restricting the $\kk$ modes to $k < (2 \pi N_{\rm g})/(3 L)$.

\begin{figure*}
    \centering
    \includegraphics[width=1.0\textwidth]{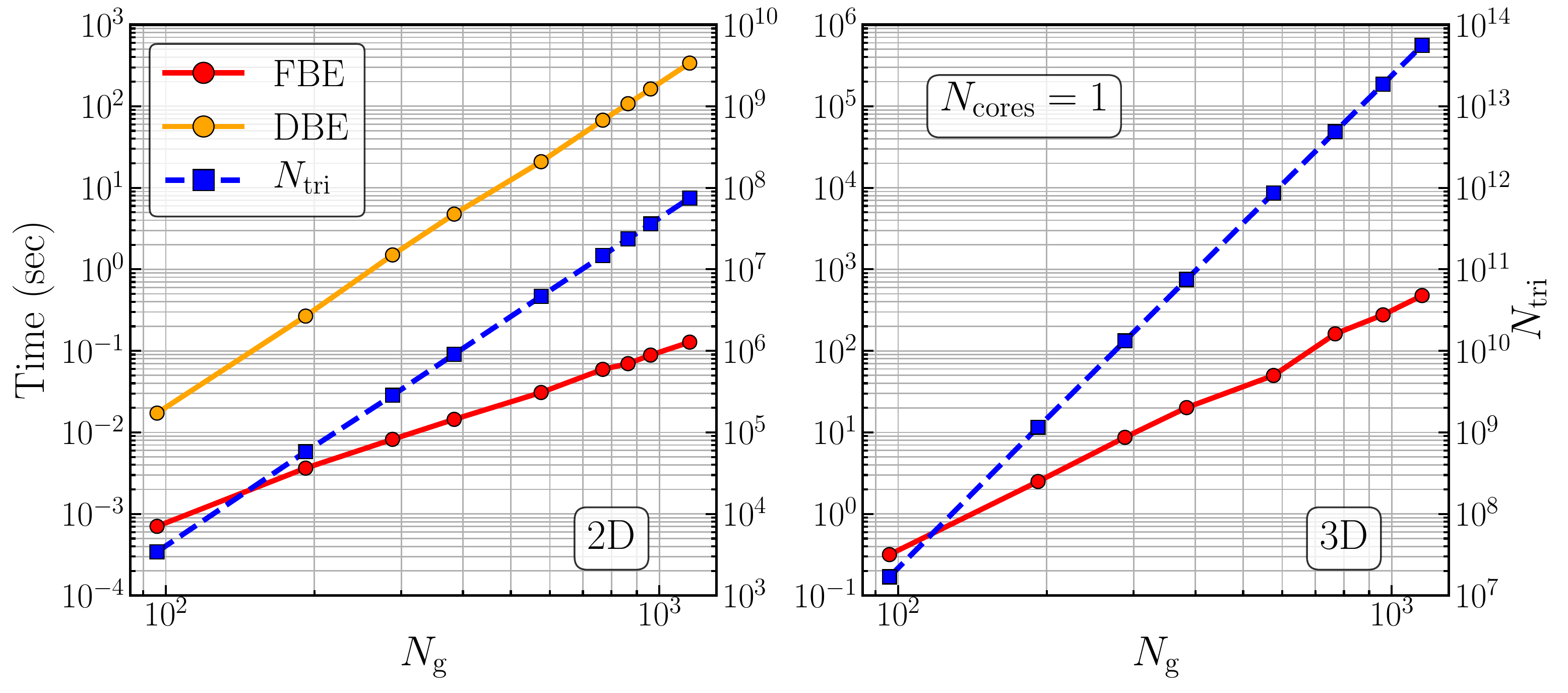}
    \caption{Shows how the computation time in bispectrum estimation varies as a function of $N_{\rm g}$ (number of grids) for FBE in both 2D (left) and 3D (right) $\kk$ space. We show the computation time for DBE for comparison, but only in 2D. Note that the quantities here are shown for the equilateral triangle bin corresponding to a single $k_1$. The computations are done on a single CPU core. The number of triangles $N_{\rm tri}$ corresponding to each value of $N_{\rm g}$ is also shown for reference.}
    \label{fig:time}
\end{figure*}

We now present a comparison of the FBE with the DBE in terms of the computational time required by  these two methods. As mentioned earlier, we expect the computational time for DBE to scale as $N_{\rm g}^6$ in In 3D. This becomes  prohibitively large as $N_{\rm g}$ is increased, and to avoid this we have carried out the comparison in 2D where the scaling is $N_{\rm g}^4$ which is more gradual. We start with the same 2D box which we had used earlier. This has a comoving length $L=215~\mpc$ and $N_{\rm g}=384$ where we have $N_r=15$ annular rings in $\kk$ space. The rings all have uniform $\delta k=0.0584~\impc$ and we have $k=0.88~\impc$ for the largest ring. Here we consider the computation time needed to calculate the bispectrum for the equilateral triangle bin where all the three modes $(\kk_{a_1}, \kk_{a_2}, \kk_{a_3})$ belong to the largest ring. In order to study how the computation time depends on $N_{\rm g}$, we have varied $N_{\rm g}$ keeping the spatial grid resolution fixed. As a consequence, the number of $\kk$ modes in each ring scales as $N_{\rm g}^2$ and the number of triangles $\Nt$ in the equilateral bin scales as $N_{\rm g}^4$. The computation time is also expected to scale as $N_{\rm g}^4$ for the DBE. The left panel of Fig. \ref{fig:time} shows the actual computation time for the DBE method as a function of $N_{\rm g}$. As expected, this scales the same as $\Nt$ (also shown in the figure), both being proportional to $N_{\rm g}^4$. In contrast, we find that the computation time for FBE scales as $\sim N_{\rm g}^2 \, \log{N_{\rm g}^2}$ which is the scaling of the FFT. The right panel of Fig. \ref{fig:time} shows how the computation time for the FBE scales with $N_{\rm g}$ in 3D, the number of triangles $\Nt$ which scales as $N_{\rm g}^6$ is shown for reference. Note that the computation time for DBE is expected to have the same $N_{\rm g}$ dependence as $\Nt$, however we have not explicitly determined this in 3D. Here we find that the computation time for the FBE scales as $\sim N_{\rm g}^3 \, \log{N_{\rm g}^3}$ which corresponds to the scaling of the FFT which is the most intensive computational step involved in the FBE.

\section{Validating the Estimator}
\label{sec:valid}
We have validated the FBE using a non-Gaussian random field $\delta(\x)$ for which it is possible to analytically predict the bispectrum. We start with a Gaussian random field $\delta_{\rm G}(\x)$ and 
calculate  $\delta(\x)$ using 
\begin{equation}
    \delta(\x)=\delta_{\rm G}(\x)+f_{\rm NG}  [\delta_{\rm G}^2(\x)-\langle \delta_{\rm G}^2(\x)  \rangle]~
\label{eq:nG}
\end{equation}
where we can interpret this as introducing non-Gaussianity through a local quadratic bias in the density field. Here the non-Gaussianity parameter $f_{\rm NG}$ controls the level of non-Gaussianity in the resulting field $\delta(\x)$. The bispectrum for this model is predicted to be 
\begin{equation}
B_{\rm Ana}(k_1,k_2,k_3)= 2 f_{\rm NG} [P(k_1)P(k_2)+P(k_2)P(k_3)+P(k_3)P(k_1)]~.
\label{eq:bana}
\end{equation} 
where we have retained only the terms that are linear in $f_{\rm NG}$. We expect  eq.~(\ref{eq:bana}) to hold for  sufficiently small values of $f_{\rm NG}$ for which $f_{\rm NG} \,  \sigma_{\rm G} \ll 1$ where $\sigma^2_{\rm G} = \langle \delta^2_{\rm G} \rangle$ is the variance of $\delta_{\rm G}$.

\begin{figure}
    \centering
    \includegraphics[scale=0.6]{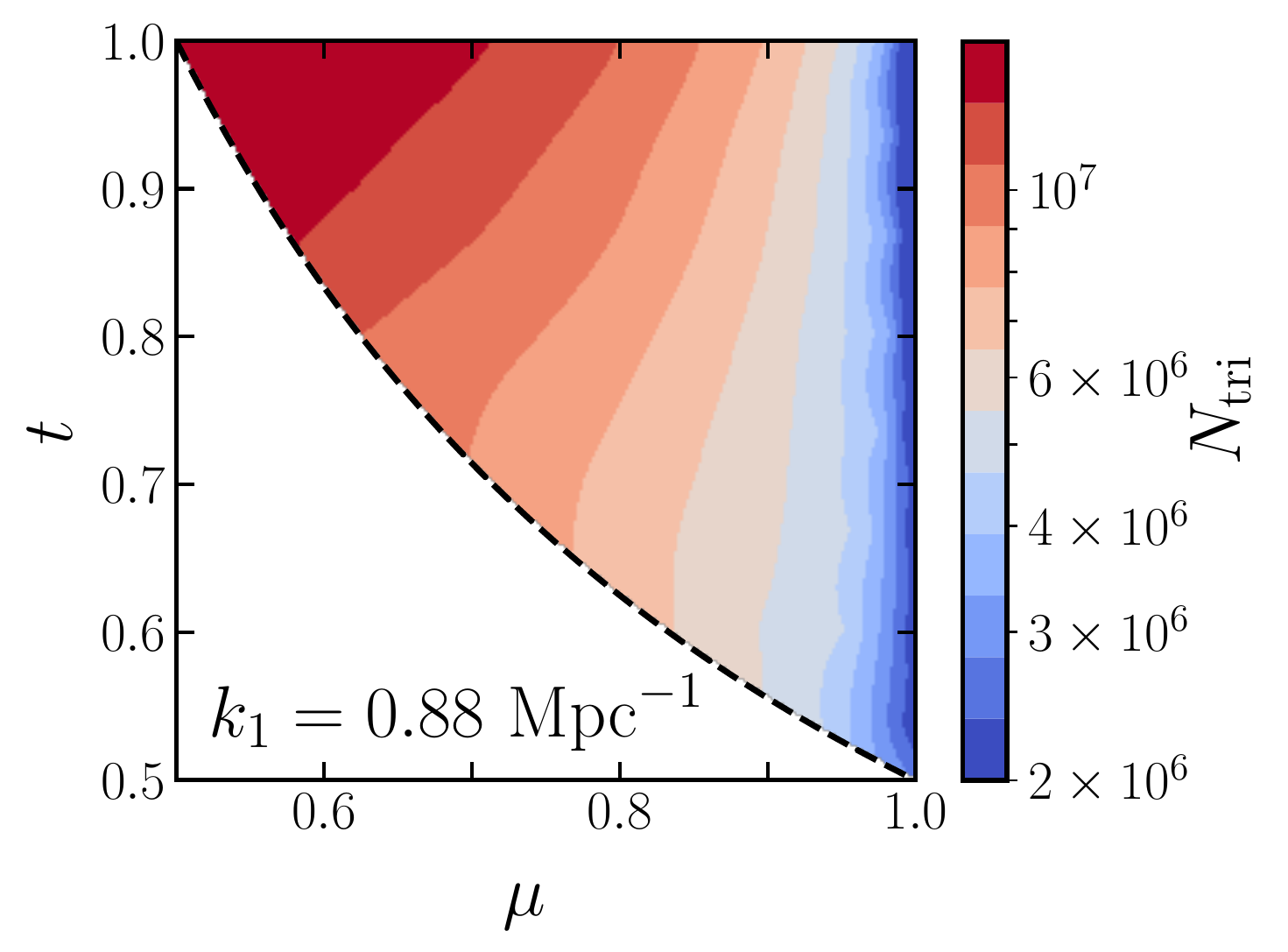}
    \caption{Shows the number of triangles ($N_{\rm tri}$) for  each of  the triangle bins shown in Fig. \ref{fig:sampling}. We have interpolated the values for a better visual representation of the results.}
    \label{fig:binum_3d}
\end{figure}

We have simulated $\delta(x)$ using a 3D box with $N_{\rm g}=384$ and $L=215~\mpc$, same as in Section \ref{subsec:estim}. We have used an input power-law power spectrum $P(k)=k^{-2}$ to simulate $\delta_{\rm G}(\x)$, and we have used $f_{\rm NG}=0.5$ to generate the non-Gaussian field $\delta(x)$ where $f_{\rm NG} \,  \sigma_{\rm G} \approx 0.2$.  We have checked that the power spectrum of the simulated non-Gaussian field does not exhibit any noticeable deviations from the input power spectrum. We have estimated the bispectrum using the same spherical shells as those considered in Section \ref{subsec:estim}. The results below are all shown for $k_1=0.88~\impc$ for which the sampling of $(\mu,t)$ space has been already shown in Fig. \ref{fig:sampling}. Fig. \ref{fig:binum_3d} shows $\Nt$ calculated using eq.~(\ref{eq:Ntri2}) for each bin. We see that $\Nt$ is largest $(\sim 10^7)$ for the equilateral triangles. The value of $\Nt$ decreases monotonically as  the shape of the triangle is deformed away from the equilateral triangle. We find that $\Nt$ falls by a factor of $\sim 2$ and $\sim 10$ respectively for right-angled triangles and linear triangles (the squeezed limit included). Considering the same shells in $\kk$ space, the value of $\Nt$ in each bin will fall if we consider smaller values of $k_1$. As discussed earlier, we expect $\Nt$ to increase as $N_{\rm g}^6$ if the value of $N_{\rm g}$ is increased.   

\begin{figure*}
    \centering
    \includegraphics[scale=0.54]{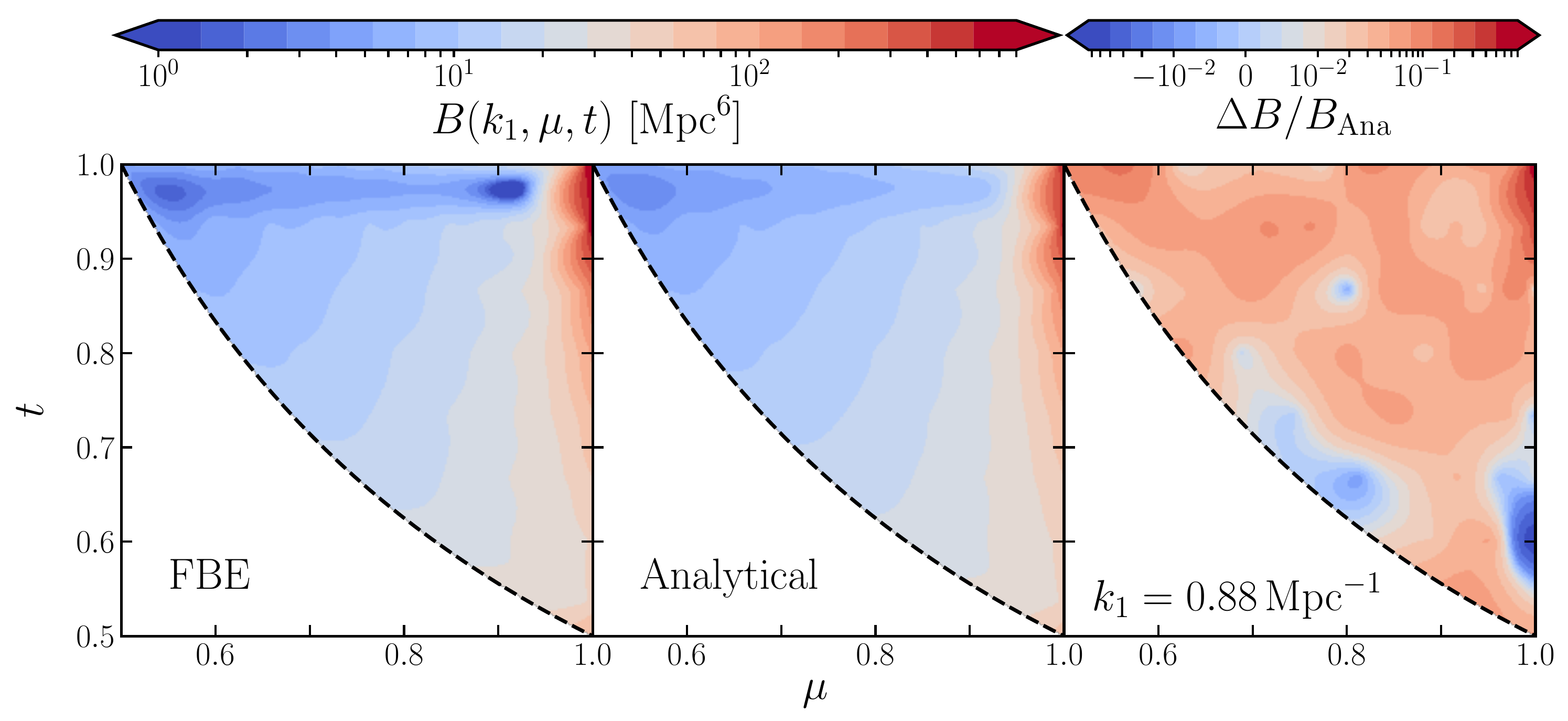}
    \caption{Shows the mean bispectrum estimated using the FBE (left panel), the analytical predictions (middle panel) and their relative deviation $\Delta B/B_{\rm Ana}=(B-B_{\rm Ana})/B_{\rm Ana}$ (right panel) for the triangle bins shown in Fig. \ref{fig:sampling}. We have interpolated the values for a visual representation of the results.}
    \label{fig:bispec_3d}
\end{figure*}

The left panel of Fig. \ref{fig:bispec_3d} shows the bispectrum estimated from the simulated $\delta(\x)$ using the FBE. Here we have considered $N_{\rm rel}=100$ independent realizations of the non-Gaussian random filed $\delta(x)$, and we show the mean estimated bispectrum here. These independent realizations were also used to estimate the variance $\sigma_B^2$ considered later. For reference, we also show the analytical predictions (eq.~\ref{eq:bana}) in the middle panel of Fig. \ref{fig:bispec_3d}. We see that the pattern of the estimated bispectrum is very similar to the analytical prediction qualitatively indicating that they are in good agreement. The bispectrum has the smallest value for equilateral triangles ($\mu=0.5,t=1$), and it increases as the triangle is deformed towards linear  triangles $(\mu=1)$. Compared to the equilateral triangle, the values of the bispectrum are nearly two orders of magnitude larger for the linear triangles. The bispectrum has the maximum value close to the squeezed limit $(\mu \rightarrow 1,t \rightarrow 1)$ along the $\mu=1$ boundary. This corresponds to the situation where $\kk_1,\kk_2,\kk_3$ are co-linear with $k_3 \ll k_2 \approx k_1$.

We have quantified the deviation between the mean estimated bispectrum $B$ and the analytical predictions $B_{\rm Ana}$ (eq.~\ref{eq:bana}) using $\Delta B/B_{\rm Ana}=(B - B_{\rm Ana})/B_{\rm Ana}$ which is shown in the right panel of Fig. \ref{fig:bispec_3d}. We note that $\Delta B/B_{\rm Ana}$ is be predominantly positive and only a few bins on the $(\mu,t)$ plane have negative values. The maximum deviation $\Delta B/B_{\rm Ana} \approx 80 \%$ occurs at $(\mu=0.997, t=1)$ which is very close to the squeezed limit, and is the top right-most bin of Fig.~\ref{fig:sampling}. We find three more bins located very close to the squeezed limit $(\mu \geq 0.99, t \geq 0.933)$ where $34 \% > \Delta B/B_{\rm Ana} > 10\%$, and one bin where $\Delta B/B_{\rm Ana} \approx 15 \%$ located at $(\mu,t)=(0.56, 1)$ which is near the equilateral limit. We have $|\Delta B/B_{\rm Ana}| \lesssim 10\%$ over the rest of $(\mu,t)$ space. The extreme negative deviation is found to be $-6.5\%$ at $(\mu,t)=(1,0.6)$, which as we discuss later, is consistent with the expected statistical fluctuations.

\begin{figure*}
    \centering
    \includegraphics[scale=0.55]{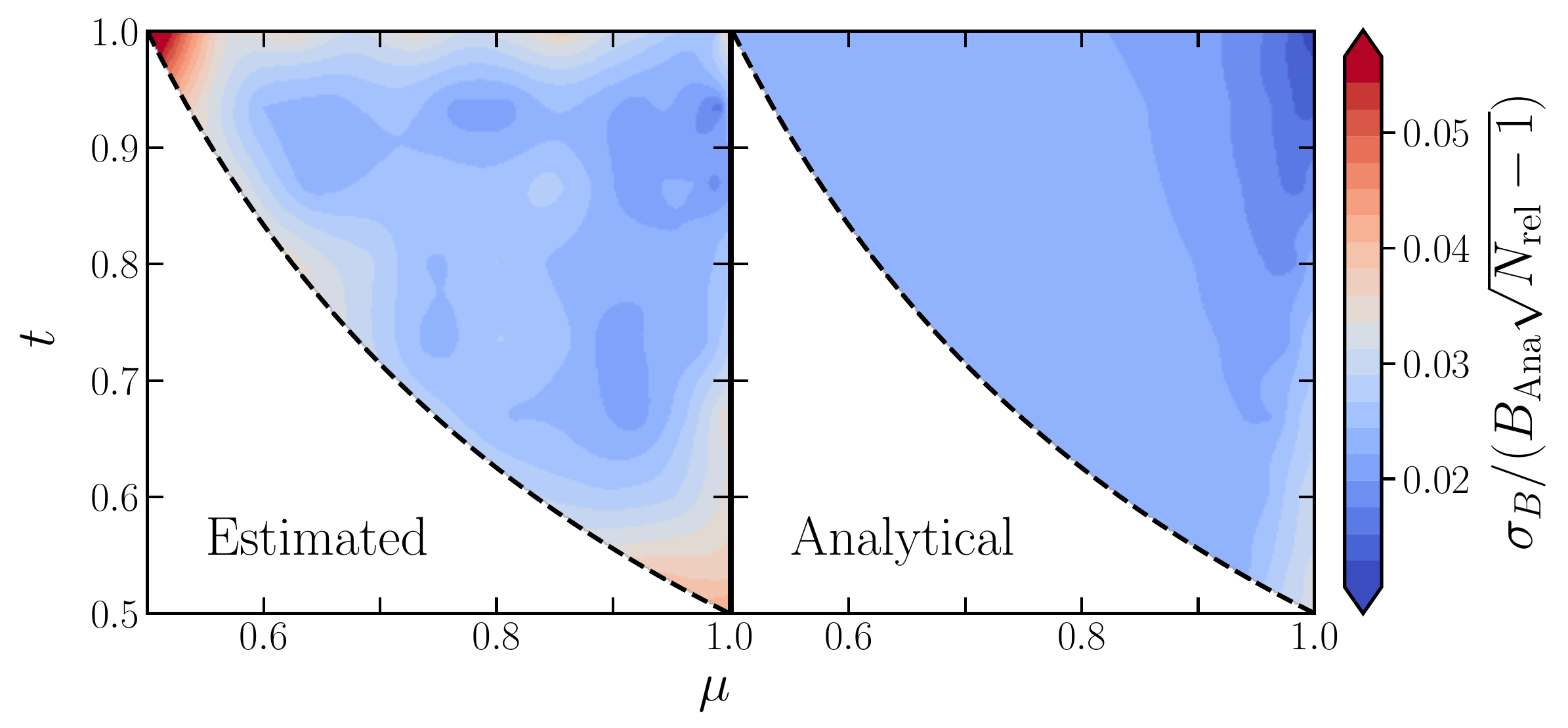}
    \caption{Shows the statistical fluctuations $\sigma_B/(B_{\rm Ana} \sqrt{N_{\rm rel}-1})$ expected for the $\Delta B/B_{\rm Ana}$ values shown in right panel of Fig. \ref{fig:bispec_3d}. The left and the right panels show the results from the simulations and the analytical predictions respectively. We have interpolated the values for a visual representation of the results.}
    \label{fig:bisig_3d}
\end{figure*}

As mentioned earlier, we have used the $N_{\rm rel}=100$ realizations to estimate $\sigma_B^2$ the variance of the estimated bispectrum. The quantity $\sigma_B/(B_{\rm Ana} \sqrt{N_{\rm rel}-1})$ shown in the left panel of Fig. \ref{fig:bisig_3d} provides an estimate of the r.m.s. statistical fluctuations expected in $\Delta B/B_{\rm Ana}$. For reference the right panel of Fig. \ref{fig:bisig_3d} shows the same quantity calculated using the analytical expression 
\begin{equation}
    \sigma^2_{B} =\frac{1}{\Nt}[{V P(k_1)P(k_2)P(k_3) + 3 B^2(k_1,k_2,k_3)}]~,
\label{eq:cv}
\end{equation}
which ignores the contributions from the higher-order terms (trispectrum etc.).
Considering the left panel of  Fig. \ref{fig:bisig_3d}, we find  that $\sigma_B/(B_{\rm Ana} \sqrt{N_{\rm rel}-1})$ has values in a rather narrow range covering $1.5\%$ to $6.7\%$. We find that the maximum value of $6.7\%$ occurs near the equilateral limit. This bin actually contains the largest number of triangles ($\Nt$, Fig. \ref{fig:binum_3d}), and we expect the sample variance to be small for this bin. The large value of $\sigma_B/(B_{\rm Ana} \sqrt{N_{\rm rel}-1})$ here is most probably due to the small value of the bispectrum $B_{\rm Ana}$ for this bin (Fig. \ref{fig:bispec_3d}). We also have somewhat large values $3.4 \% <\sigma_B/(B_{\rm Ana} \sqrt{N_{\rm rel}-1})< 3.9\%$ near the stretched limit, and also along the top and lower boundaries which correspond to the L and S isosceles triangles respectively.  We have $\sigma_B/(B_{\rm Ana} \sqrt{N_{\rm rel}-1}) \lesssim 3 \%$ across  most  of the remaining $(\mu,t)$ space, with the lowest value $1.6\%$ occurring near the squeezed limit which is also where the bispectrum has the largest value. Comparing the two panels, we see that the values of $\sigma_B/(B_{\rm Ana} \sqrt{N_{\rm rel}-1})$ cover an even smaller range ($1.1\%$ to $3.2\%$) in the right panel. The values in the two panels are quite similar, with the difference that the large value seen for the equilateral limit in the left panel is not present in the right panel. Our analysis shows that the error estimates from $100$ realizations of the simulations are consistent with analytical predictions. We shall use the results from the left panel for the subsequent discussion. 

We have compared the values of $\Delta B/B_{\rm Ana}$ (right panel of Fig. \ref{fig:bispec_3d}) with $\sigma_B/(B_{\rm Ana} \sqrt{N_{\rm rel}-1})$ (left panel of Fig. \ref{fig:bisig_3d}) in order to determine whether the deviations between the estimated bispectrum and the analytical predictions are due to statistical fluctuations or if the estimator systematically underestimates or overestimates the bispectrum. We recollect that we have $|\Delta B/B_{\rm Ana}| \lesssim 10\%$ across most of $(\mu,t)$ space. We find that $|\Delta B/B_{\rm Ana}| \le 3 \sigma_B/(B_{\rm Ana} \sqrt{N_{\rm rel}-1})$ for most of the bins which satisfy this condition. This indicates that the deviations are consistent with statistical fluctuations for most of the bins. However, we find that for a few bins near the squeezed limit and a single bin near the equilateral limit $\Delta B/B_{\rm Ana}$ has values $> 10 \%$ which is considerably in excess of the expected statistical fluctuations. This indicates that the FBE overestimates the bispectrum at these bins. Note that the bispectrum $B(k_1,\mu,t)$ varies relatively rapidly as a function of $(\mu,t)$ near the squeezed and the equilateral limits, and the deviations here are possibly due to the finite width of the $(\mu,t)$ bins.

To test this, we have redone the entire analysis using $N_r=30$ shells each  having a smaller width of $\delta k=0.0292~\impc$ which corresponds to an unit grid spacing. We now find that the largest deviation $\Delta B/B_{\rm Ana} = 22.3\%$ occurs at $(\mu,t)=(0.53,1)$ which is near the equilateral limit. This value is nearly $4$ times smaller than the value of the largest deviation in the earlier analysis where we had used shells with  $\delta k=0.0584~\impc$. Recall that earlier we had the largest deviation of $\Delta B/B_{\rm Ana} =80\%$ in the bin which is closest to the squeezed limit. We now have $\Delta B/B_{\rm Ana} =-7.8\%$ and $-5\%$ at $(\mu,t)=(0.9992,1)$ and $(0.997, 1)$ respectively which are the two bins closest to the squeezed limits. This clearly demonstrates that the large relative errors noted earlier near the squeezed and equilateral limits occurs due to the finite bin width, and these can be reduced by choosing smaller bins. However, it is necessary to note that the predicted statistical errors (eq. \ref{eq:cv}) increase if we reduce the bin size. A judicious choice of the bins would depend on the particular context where the estimator is being used. Here we have used a relatively simple choice for the purpose of demonstrating and validating the estimator.

\begin{figure}
    \centering
    \includegraphics[scale=0.65]{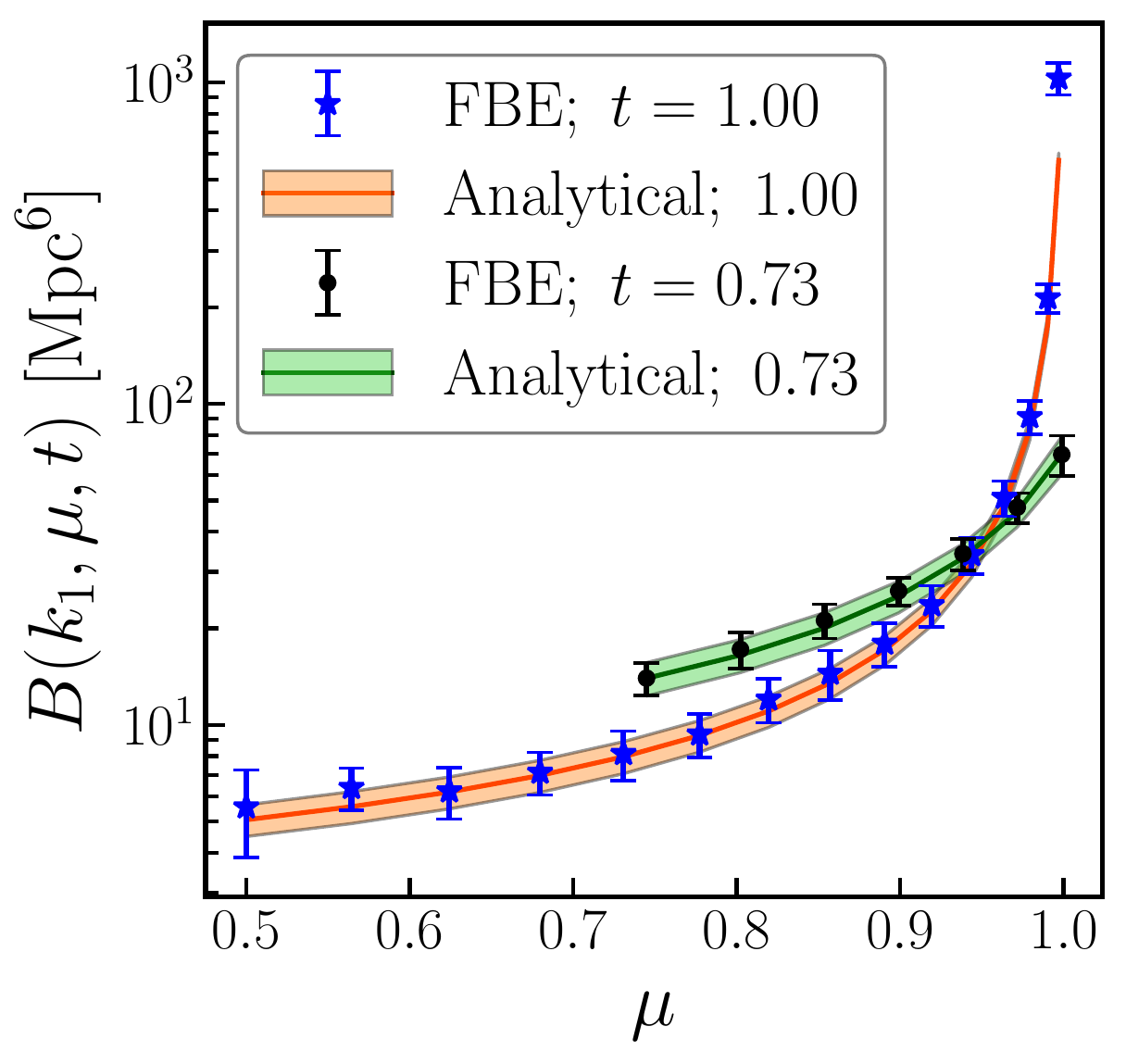}
    \caption{Shows $B(k_1,\mu,t)$ as a function of $\mu$ for $t=1.00$ and $0.73$ with $k_1=0.88~\impc$. The black and blue points show the mean bispectrum estimated from $N_{\rm rel}=100$ realizations of the simulations and the  $5\sigma$ error-bars denote the statistical fluctuations estimated from the simulations. The solid lines shows the analytical prediction for the bispectrum while the shaded region shows the analytical predictions for the $5\sigma$ statistical fluctuations.}
    \label{fig:bispec_line}
\end{figure}

For a more direct comparison between the estimated $B$ and the analytical prediction, Fig. \ref{fig:bispec_line} shows the values of these quantities as a function of $\mu$ along two sections of the $(\mu,t)$ plane which respectively correspond  to $t=1.00$ and $0.73$. The figure also shows the expected statistical fluctuations $5 \times \sigma_B/\sqrt{N_{\rm rel}-1}$, with the value estimated from the simulations as well as the analytical predictions both being shown here. Note that the expected statistical fluctuations are rather small, and we have shown $5$ times the expected fluctuations to make these visible. We see that the values of $B$ increases by at least one order of magnitude as $\mu$ increases and approaches $\mu=1$ (linear triangles). Further, we find that the expected statistical fluctuations get relatively smaller as $\mu \rightarrow 1$. Considering $t=0.73$ first, we find that for all values of $\mu$ the estimated bispectrum is in very good agreement with the analytical predictions, the deviations between the two being well within the expected $5 \sigma$ fluctuations. We also find that the estimated statistical fluctuations are in good agreement with the analytical predictions. Considering $t=1$, we see that values of the estimated bispectrum are in good agreement with the analytical predictions for all values of $\mu$ except for a single bin which corresponds to the squeezed limit $(\mu \rightarrow 1, t=1)$ where the estimated bispectrum is roughly twice the analytical predictions. Further, the estimated statistical fluctuations are found to be roughly consistent with the analytical predictions for $\mu \le 0.9$. The estimated values exceed the analytical predictions for larger values of $\mu$. As discussed earlier, the deviations near the squeezed limit possibly arise due to the fact that the value of the bispectrum $B(k_1,\mu,t)$ varies very rapidly as a function of $(\mu,t)$ in this region of $(\mu,t)$ space. Note that this region of parameter space has the smallest values of $\Nt$. The small number of triangles in the bin possibly also contributes to the large deviations. 


\subsection{Linear redshift space distortion}\label{subsec:rsd}

\begin{figure*}
    \centering
    \includegraphics[scale=0.54]{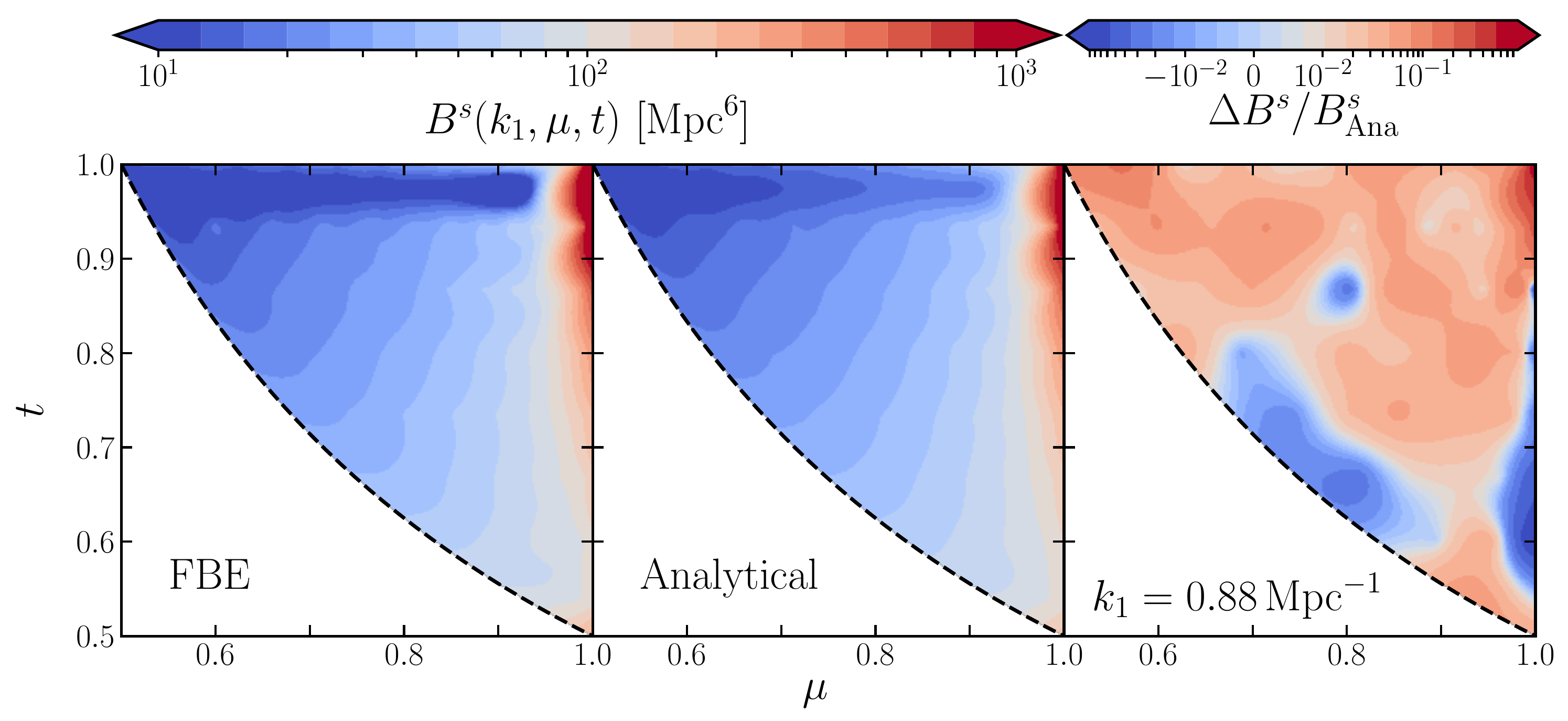}
    \caption{Shows the plot same as in Fig. \ref{fig:bispec_3d} but for linear RSD.}
    \label{fig:bispec_3d_RSD}
\end{figure*}

The redshift space distortion (RSD), which arises due to the peculiar velocities, is inevitable in several observations such as galaxy-redshift surveys and \HI intensity mapping etc. It is possible to quantify the redshift space bispectrum by decomposing it into multipole moments $\bar{B}^m_\ell(k_1,\mu,t)$ using spherical harmonics \cite{Bharadwaj_2020}. In this sub-section we have applied linear RSD to the simulated non-Gaussian random field (eq.~\ref{eq:nG}) for which the analytical predictions for all the non-zero multipole moments $\bar{B}^m_\ell(k_1,\mu,t)$ are available in \cite{Bharadwaj_2020}. Here we have validated the FBE by using it to estimate the monopole $\bar{B}^0_0(k_1,\mu,t)$ of the redshift space bispectrum and comparing this with the analytical prediction.  We note that the expression for the analytical prediction is rather lengthy and we have not shown it here, the reader is referred to  eqs. (24) and (26) of \cite{Bharadwaj_2020} for details. It is important to note that the analytical predictions presented in \cite{Bharadwaj_2020} considers the flat-sky approximation. This is not a good approximation for the future large surveys and we defer it to our future studies. Further, since the entire analysis here is restricted to the monopole we use the  notation $B^s(k_1,\mu,t) \equiv \bar{B}^0_0(k_1,\mu,t)$ throughout the subsequent discussion.

Considering $\Delta(\kk)$ the Fourier transform of the the non-Gaussian filed $\delta(x)$ (eq.~\ref{eq:nG}), we introduce the effect of linear RSD \cite{Hamilton_1998} using 
\begin{equation}
    \Delta^s(\kk)=(1+\beta_1 \mu_1^2)\Delta^r(\kk)~. 
    \label{eq:RSD}
\end{equation}
where $\Delta^s(\kk)$ is the redshift space counterpart of $\Delta(\kk)$, $\beta_1$ is the linear RSD parameter and $\mu_1=(\hat{z}\cdot \kk)/|\kk|$ is the cosine of the angle between $\kk$ and LoS direction $\hat{z}$. Here we use $\beta_1=1$ throughout. As mentioned earlier, we have used the FBE to estimate  $B^s(k_1,\mu,t)$ the monopole of the redshift space bispectrum for this non-Gaussian random field.


The left and middle panels in Fig. \ref{fig:bispec_3d_RSD} show the shape dependence of the redshift space bispectrum for $k_1=0.88~\impc$. We find that the shape dependence of the estimated bispectrum $B^s(k_1,\mu,t)$ (left panel) is qualitatively very similar to the analytical predictions $B^s_{\rm Ana}(k_1,\mu,t)$ (middle panel). We also find the shape dependence of $B^s(k_1,\mu,t)$ is qualitatively very similar to that of its real-space counterpart $B(k_1,\mu,t)$ (Fig. \ref{fig:bispec_3d}), except that $B^s(k_1,\mu,t)$ has larger amplitude. The corresponding enhancement factor $B^s/B$ depends on $(\mu,t)$ which quantify the shape of the triangle. We find that the  estimated values of $B^s/B$ are  in good agreement with the analytical predictions \cite[see Fig. 3 of][]{Bharadwaj_2020}, and we do not explicitly show this here. The deviations between the estimated and analytically predicted enhancement factors are roughly within $\pm 7\%$ for all the bins in $(\mu,t)$ plane.

We have quantified the deviation between the mean estimated bispectrum $B^s$ and the analytical predictions $B^s_{\rm Ana}$ using $\Delta B^s/B^s_{\rm Ana}=(B^s - B^s_{\rm Ana})/B^s_{\rm Ana}$ which is shown in the right panel of Fig. \ref{fig:bispec_3d_RSD}. Like $\Delta B/B_{\rm Ana}$, we find that $\Delta B^s/B^s_{\rm Ana}$ also is  predominantly positive and only a few bins on the $(\mu,t)$ plane have negative values. The maximum deviation $\Delta B^s/B^s_{\rm Ana} \approx 89 \%$ occurs at $(\mu=0.997, t=1)$ which is very close to the squeezed limit, and we have three more bins located very close to the squeezed limit $(\mu \geq 0.99, t \geq 0.933)$ where $30 \% > \Delta B^s/B^s_{\rm Ana} > 15\%$, and one bin where $\Delta B^s/B^s_{\rm Ana} \approx 14 \%$ located at $(\mu,t)=(0.56, 1)$ which is near the equilateral limit. We have $|\Delta B/B_{\rm Ana}| \lesssim 10\%$ over the rest of $(\mu,t)$ space. The extreme negative deviation is found to be $-9.3\%$ at $(\mu,t)=(1,0.6)$, which we consider to be consistent with the expected statistical fluctuations. Overall, the deviations $\Delta B^s/B^s_{\rm Ana}$ are very similar to $\Delta B/B_{\rm Ana}$ shown in Fig.~\ref{fig:bisig_3d}. 

\begin{figure}
    \centering
    \includegraphics[scale=0.65]{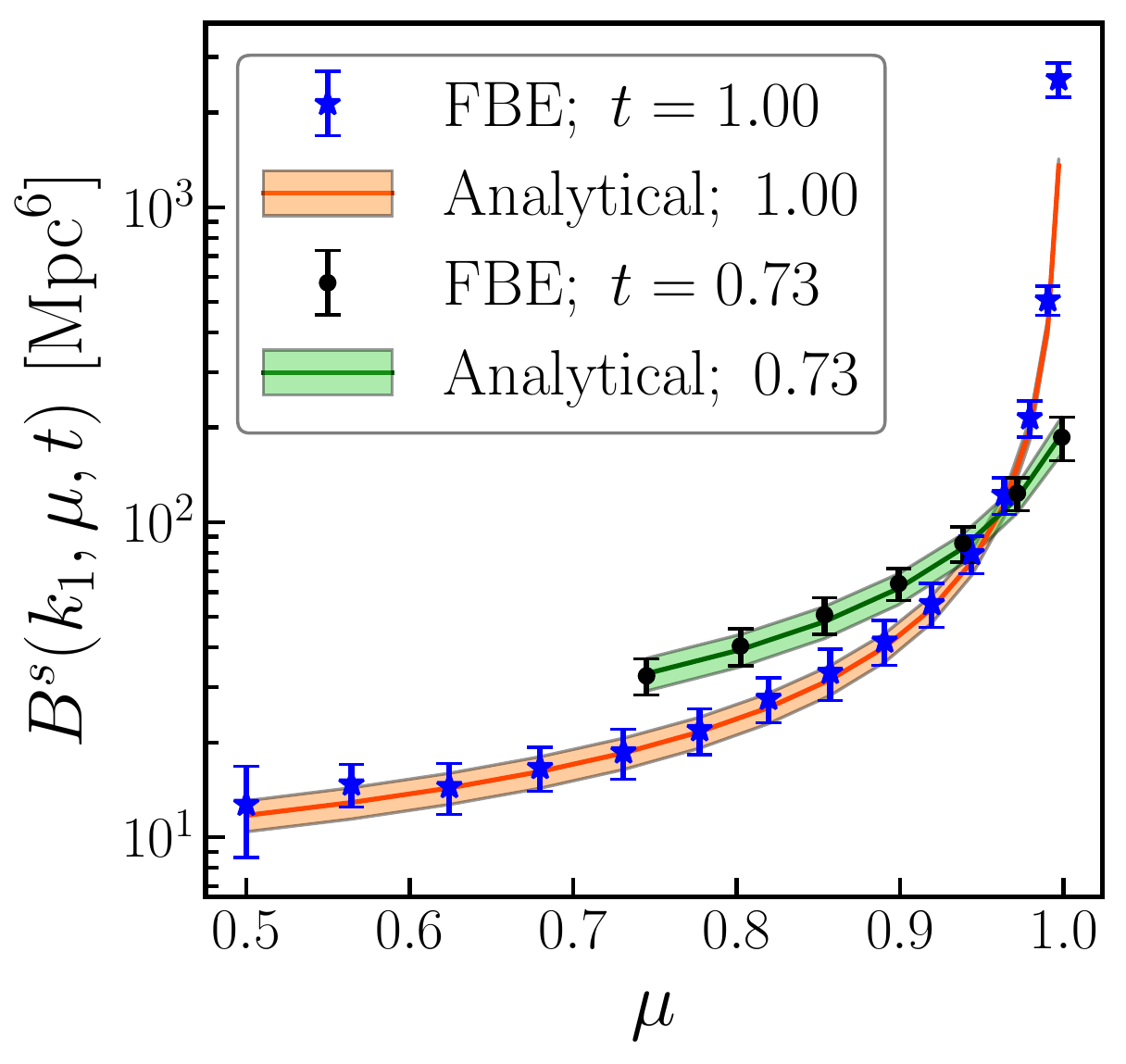}
    \caption{Shows the plot same as in Fig. \ref{fig:bispec_line} but for linear RSD.}
    \label{fig:bispec_line_RSD}
\end{figure}

Fig. \ref{fig:bispec_line_RSD} shows a comparison between $B^s$ and $B^s_{\rm Ana}$ as a function of $\mu$ along two sections of the $(\mu,t)$ plane which respectively correspond to $t=1.00$ and $0.73$. Similar to Fig. \ref{fig:bispec_line}, this also shows the expected $5 \sigma$ statistical fluctuations from the simulations as well as the analytical predictions. $B^s$ and $B^s_{\rm Ana}$ are consistent well within $5 \sigma$ fluctuations for both $t=1$ and $0.73$, except for one bin closest to the squeezed limit $(\mu \rightarrow 1, t=1)$. The estimated bispectrum for this bin is larger than the corresponding analytical prediction by a factor slightly less than $2$. We find that the $\mu$ dependence of the redshift space bispectrum is qualitatively very similar to the corresponding real space estimates shown in Fig. \ref{fig:bispec_line}. As mentioned, $B^s$ and $B$ are related through an enhancement factor due to RSD. This enhancement factor is minimum for the equilateral triangle where $B^s/B \approx 2.29$. The enhancement factor increases as the value of $\mu$ increases. Considering $t=1$, we have $B^s/B \approx 2.47$ at $\mu \rightarrow 1$. Considering $t=0.73$, we find that $B^s/B$ has values $2.32$ and $2.68$ respectively at $\mu \approx 0.74$ and $1$.

In summary, considering both real and redshift space we find that the estimated bispectrum is in very good agreement with the analytical predictions over nearly the entire $(\mu,t)$ space, barring a few bins near the squeezed limit and a single bin near the equilateral limit. 


\section{Summary and Discussion}
\label{sec:dis}

The bispectrum, which is a function of the triangles formed by three wave vectors, is the lowest order statistics which is sensitive to non-Gaussianity. Its dependence on the shape and size of the triangle contains a wealth of cosmological information. We use $k_1$, the length of the largest side of the triangle, to parameterize the size dependence and we use two dimensionless parameters $(\mu,t)$ to parameterize the shape dependence. This parameterization allows us to disentangle the shape dependence from the size dependence and study these separately.

In this paper, we consider an estimator for the bin-averaged bispectrum $B(k_1,\mu,t)$ where $k_1$ and $(\mu,t)$ respectively refer to the average size and shape of the triangles in the bin. Considering a density field defined on a grid covering a finite volume with periodic boundary conditions, we use spherical shells of uniform radial extent $\delta k$ in $\kk$ space. Each combination of three shells corresponds to a bin of triangles with a different set of $(k_1, \mu,t)$ values. The number of triangles $\Nt$ in each bin scales as $N_{\rm g}^6$ where $N_{\rm g}$ is the number of grid points along each side of the volume. Considering a straightforward implementation that estimates the bispectrum by directly looping through the $\kk$ modes in two of the shells and checks whether the $\kk$ mode required to close the triangle lies within the third shell (Direct Bispectrum Estimator, DBE), the computation time scales as $\Nt \propto N_{\rm g}^6$. This computation time scales very steeply as $N_{\rm g}$ is increased, and DBE becomes extremely computation intensive for large $N_{\rm g}$. To overcome this, here we have implemented an FFT based \cite{Sefusatti_thesis, Jeong_thesis} fast estimator (Fast Bispectrum Estimator, FBE) where the computation time scales as $N_{\rm g}^3 \, \log{N_{\rm g}^3}$ which is much less steep. To give an example, we find that for $N_{\rm g}=10^2$ the FBE requires $0.3$ sec to estimate the bispectrum at a particular bin for which DBE requires $150$ sec. Considering $N_{\rm g}=10^3$ instead, we find that the computation time is $300$ sec for FBE whereas it is expected to scale to $1.5\times 10^8$ sec for the DBE. We see that the FBE provides a tremendous advantage when the grid size is large ($N_{\rm g} \sim 10^3$ and larger). However, as we explicitly demonstrate here, the FFT based estimator has its own limitation which arises due to the periodic boundary condition in Fourier space. We avoid this by restricting the largest $\kk$ mode included in the analysis to $k < (2 \pi N_{\rm g})/(3 L)$. We note that a similar consideration is also expected to hold for an FFT based $p$-th order polyspectrum estimator for which it will be necessary to restrict the largest $\kk$ mode to $k < (2 \pi N_{\rm g})/(p L)$.

We have validated the FBE by applying it to a non-Gaussian random field (eq. \ref{eq:nG}) for which the expected bispectrum can be analytically calculated ($B_{\rm Ana}$,  eq. \ref{eq:bana}). We have used $N_{\rm rel}=100$ realizations to calculate the mean $B$ and the r.m.s. fluctuations $\sigma_B$ of the estimated bispectrum. We find that the shape dependence of $B$ and $B_{\rm Ana}$ are qualitatively very similar (Fig. \ref{fig:bispec_3d}). We find that for most of the bins on $(\mu,t)$ plane, the fractional deviation $(B-B_{\rm Ana})/B_{\rm Ana}$ is within $\pm 10 \%$ which is well within the $3 \sigma$ statistical fluctuations. However, there are a few bins near the squeezed limit and one bin near the equilateral limit where the fractional deviations are larger $(> 10 \%)$. We have the largest fractional deviation of $80\%$ at the bin $(\mu =0.997, t=1)$ very close to the squeezed limit. We attribute this to the fact that the bispectrum varies very rapidly here, and this results in a substantial change in the value of the bispectrum across the finite extent of the bin.

Redshift space distortion (RSD) is an important effect which introduces an anisotropy along the line-of-sight (LoS) direction. The RSD anisotropy causes the bispectrum to depend on the orientation of the triangle $(\kk_1,\kk_2,\kk_3)$ with respect to the LoS direction. We can quantify this anisotropy by decomposing the redshift space bispectrum into multipole moments, however it is necessary to ensure that the estimator uniformly samples all possible triangle orientations so as to correctly estimate the various multipole moments. Here we have used the FBE to estimate $B^s(k_1,\mu,t)$ the monopole component of the redshift space bispectrum. In order to validate this, we have applied linear RSD to the non-Gaussian random field, a situation for which the expected multipole moments of the bispectrum can be analytically calculated. We find that the estimated $B^s(k_1,\mu,t)$ values are qualitatively very similar to the analytical predictions $B_{\rm Ana}$. Further, the fractional deviations $(B^s-B^s_{\rm Ana})/B^s_{\rm Ana}$ are quantitatively very similar to those for the real space counterpart $(B-B_{\rm Ana})/B_{\rm Ana}$.  

In conclusion, we have validated that the FBE provides a fast and reasonably accurate estimate of the bispectrum, both with and without the RSD effect.  

\acknowledgments
We acknowledge National Supercomputing Mission (NSM) for providing computing resources of `PARAM Shakti' at IIT Kharagpur, which is implemented by C-DAC and supported by the Ministry of Electronics and Information Technology (MeitY) and Department of Science and Technology (DST), Government of India. We also acknowledge the computing facility at the Centre for Theoretical Studies at IIT Kharagpur which has been extensively used for the initial development. The authors would like to thank Sk. Saiyad Ali for his comments. AKS would like to thank Anjan Kumar Sarkar for fruitful discussions. DS acknowledges support from the Azrieli Foundation for his Postdoctoral Fellowship.

\bibliographystyle{JHEP_new} 
\bibliography{refbs}
\end{document}